# UNIVERSAL QUANTUM GATES FOR SINGLE COOPER PAIR BOX BASED QUANTUM COMPUTING

P. ECHTERNACH[1], C. P. WILLIAMS[2], S.C. DULTZ[1], P. DELSING[3], S. BRAUNSTEIN[4], AND J. P. DOWLING[2]

[1]*Microdevices Laboratory, Jet Propulsion Laboratory, California Institute of Technology, Pasadena, California*

[2]*Quantum Computing Group, Jet Propulsion Laboratory, California Institute of Technology, Pasadena, California*

[3]*Chalmers University of Technology, Göteborg, Sweden*

[4]*University of Wales, Bangor, United Kingdom*

**Introduction**

Several schemes have been proposed for implementing quantum computer hardware in solid state quantum electronics. These schemes use electric charge[1,2,3], magnetic flux[4,5,6], superconducting phase[7,8,9,10], electron spin[11,12,13,14], or nuclear spin[15,16] as the information bearing degree of freedom. Each scheme has various pros and cons, but the one based on harnessing quantized charge is especially appealing because the necessary superconducting circuitry for such a qubit can be fabricated using present day e-beam lithography equipment, and quantum coherence, essential for creating superposed and entangled states, has been demonstrated experimentally[17]. Moreover, the fidelity and leakage of such gates is understood[18]. These qualities make the SCB-based qubit a strong contender for the basic element of a quantum computer. Indeed, today s e-beam fabrication technology is sufficiently mature that it would be a simple matter to create a quantum circuit having thousands of quantum gates within a matter of a few hours! Of course, it remains to be seen whether such large-scale quantum circuits could be operated coherently en masse. Nevertheless, the relative ease of fabricating SCB-based quantum gates leads one to consider computer architectural issues related large scale SCB-based quantum circuits.

From an architectural perspective, the existing proposals for SCB-based qubits and quantum gates are sub-optimal. For example, the scheme of Sch n et al. [19] uses the time at which a gate operation begins as one of the parameters that determine the unitary operation the gate is to perform. While this is certainly allowed physically, and could even be argued to be ingeniously efficient, it is not a good decision from the perspective of building reliable and scaleable quantum computers. If the starting time is a parameter, a given quantum gate would need different implementations at different times. Moreover, as the computation progressed, timing errors would accumulate leading to worsening gate fidelity. Furthermore, Sch n et al. also use the duration of the gate operation as a free parameter that determines the unitary transformation the gate is to perform. Again, this is a poor decision from a computer architecture perspective, as it means that different gates would take different times making it difficult to synchronize parallel quantum gate operations in large circuits. To address both of these problems we have developed an approach to universal quantum computation in SCB-based quantum computing that specifically avoids using time as a free parameter. Instead, our gates operate by varying only voltages and magnetic fluxes in a controlled fashion.

To make a practical design for a quantum computer, one must specify how to decompose any valid quantum computation into a sequence of elementary 1- and 2-qubit quantum gates that can be realized in physical hardware that is feasible to fabricate. The set of these 1- and 2-qubit gates is arbitrary provided it is *universal*, i.e., capable of achieving any valid quantum computation from a quantum circuit comprising only gates from this set. Traditionally the set of universal gates has been taken to be the set of all 1-qubit quantum gates in conjunction with a single 2-qubit gate called controlled-NOT. However, many equally good universal gate sets exist[20] and there might be an advantage in using a non-standard universal gate set if certain gate designs happen to be easier to realize in one hardware context than another[21,22]. Certainly it has been known for some time that the simple 2-qubit exchange interaction (i.e., the SWAP gate) is as powerful as CNOT as far as computational universality is concerned. It makes sense therefore, to see what gates are easy to make and then extend them into a universal set. This is the strategy pursued in this paper. In particular, we show, in the context of SCB-based qubits, that we can implement any 1-qubit operation and a special (new) 2-qubit operation called the complex SWAP (or *iSWAP* for short). We then prove that, taken together, *iSWAP* and all 1-qubit gates is universal for quantum computation.

## 2   SCB-based Qubits

A Single Cooper Pair Box is an artificial two-level quantum system comprising a nanoscale superconducting electrode connected to a reservoir of Cooper pair charges via a Josephson junction. The logical states of the device, $|0\rangle$ and $|1\rangle$, are implemented physically as a pair of charge-number states differing by $2e$ (where $e$ is the charge of an electron). Typically, some $10^9$ Cooper pairs are involved. Transitions between the logical states are accomplished by tunneling of Cooper pairs through the Josephson junction. Although the two-level system contains a macroscopic number of charges, in the superconducting regime they behave collectively, as a Bose-Einstein condensate, allowing the two logical states to be superposed coherently. This property makes the SCB a candidate for the physical implementation of a qubit.

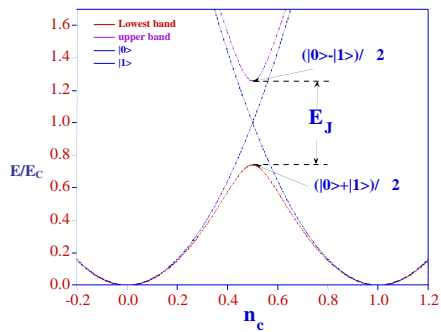

Figure 1. The level diagram for an SCB-based qubit.

The SCB-qubit gained prominence in 1999 when Nakamura et al. demonstrated coherent oscillations between the $|0\rangle$ and $|1\rangle$ states[17]. This was the first time such macroscopic coherent phenomena had been seen experimentally and distinguishes the SCB approach from other solid state schemes in which similar macroscopic coherences have not yet been demonstrated. Recently there were two experimental demonstrations that a superconducting quantum interference device (SQUID) can be put into a superposition of two-magnetic flux states[23,24], albeit those demonstrations used frequency domain experiments, instead of the more advanced tme domain experiments.

Our qubit consists of a *split* tunnel junction as this allows us to control the Josephson coupling by varying the externally applied magnetic flux according to:

$$E_J(\Phi_{ext}) = -2E_J \cos\left(\frac{\pi \Phi_{ext}}{\Phi_0}\right) \quad (1)$$

where $\Phi_0$ is the quantum of magnetic flux and $E_J^{intrinsic}$ is given by the Ambegaokar-Baratoff relation in the low temperature approximation: $E_J^{intrinsic} = \frac{h\Delta}{8e^2 R_N}$ (in which $h$, $\Delta$, and $R_N$ are Planck s constant, the superconducting energy gap and the normal tunneling resistance of the junction respectively). Figure 2 shows a schematic diagram of our qubit.

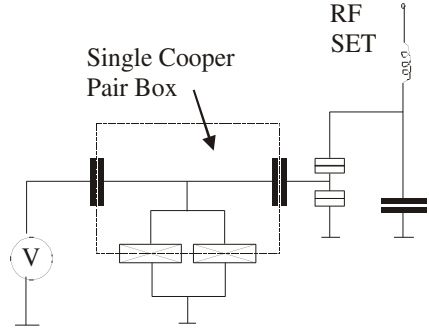

Figure 2. Schematic diagram of a single SCB-based qubit with an adjoining RF SET readout.

The Hamiltonian for the qubit is $H = 4E_C \sum_n (n - n_C)^2 - E_J(\Phi_{ext})\cos(\phi)$, where $n$ is the number of excess Cooper pairs on the island, $n_C = \frac{C_g V}{2e}$, $E_C = \frac{e^2}{2C_\Sigma}$, $C_\Sigma = C_J + C_g$ (neglecting for the moment the capacitance to the RF-SET), and $\phi$ is the difference in phase of the superconducting state across the junction. In the basis of excess Cooper pair number states, $|n\rangle$, assuming $E$, the Hamiltonian reduces to:

$$H_1 = \begin{pmatrix} -\tfrac{1}{2}E(V) & -\tfrac{1}{2}E_J(\Phi_{ext}) \\ -\tfrac{1}{2}E_J(\Phi_{ext}) & +\tfrac{1}{2}E(V) \end{pmatrix} + \begin{pmatrix} E_O & \\ & E_O \end{pmatrix} \quad (2)$$

where $E_O = E_C(n_c^2 - n_c + 1/2)$, $E(V) = E_C\left(1 - \frac{C_G V}{e}\right)$ and $E_J(\Phi_{ext}) = 2E_J^{intrinsic}\left|\cos(\frac{\pi \Phi_{ext}}{\Phi_0})\right|$. The two parameters $V$ and $\Phi_{ext}$ can be adjusted to achieve different Hamiltonians and hence different 1-qubit quantum gates.

## 3   One-Qubit Gates

The 1-qubit Hamiltonian, $H_1$, acting for a time $\Delta t$ induces a 1-qubit quantum gate operation given by:

$$U_1 = \exp\left(-\frac{iH_1 \Delta t}{\hbar}\right) \quad (3)$$

We assume that the Hamiltonian can be switched on and off quickly so that the interval $\Delta t$ is sharp. The fact that $H_1$ has a symmetric structure means that we are only able to implement a limited set of primitive unitary transformations. Nevertheless, it turns out that these primitive transformations can be composed to achieve *arbitrary* 1-qubit gates. The proof is via a factorization of an arbitrary $2 \times 2$ unitary matrix into a product of rotation matrices. Specifically, the matrix for an arbitrary 1-qubit gate is described mathematically by[25]

$$U(\alpha,\theta,\beta) = \begin{pmatrix} e^{i\left(\frac{\alpha+\beta}{2}\right)}\cos\left(\frac{\theta}{2}\right) & e^{i\left(\frac{\alpha-\beta}{2}\right)}\sin\left(\frac{\theta}{2}\right) \\ -e^{-i\left(\frac{\alpha-\beta}{2}\right)}\sin\left(\frac{\theta}{2}\right) & e^{-i\left(\frac{\alpha+\beta}{2}\right)}\cos\left(\frac{\theta}{2}\right) \end{pmatrix} \tag{4}$$

Such a matrix can be factored into the product of rotations about just the $z$- and $x$-axes.

$$U(\alpha,\theta,\beta) = R_z\left(\alpha - \frac{\pi}{2}\right) \cdot R_x(\theta) \cdot R_z\left(\beta + \frac{\pi}{2}\right) \tag{5}$$

where $R_z(\xi) = \exp(i\xi\sigma_z/2)$ is a rotation′ about the $z$-axis through angle $\xi$, $R_x(\xi) = \exp(i\xi\sigma_x/2)$ is a rotation about the $x$-axis through angle $\xi$, and $\sigma_i : i \in \{x, y, z\}$ are Pauli spin matrices $\sigma_x = \begin{pmatrix} 0 & 1 \\ 1 & 0 \end{pmatrix}$, $\sigma_y = \begin{pmatrix} 0 & -i \\ i & 0 \end{pmatrix}$, $\sigma_z = \begin{pmatrix} 1 & 0 \\ 0 & -1 \end{pmatrix}$.

It is therefore sufficient to configure the parameters in $H_1$ to perform rotations about just the $z$- and $x$-axes to achieve an arbitrary 1-qubit gate. From equations (2) and (3), we find that $R_z(\xi)$ can be achieved within time $\Delta t$ by setting $\Phi_{ext} = \Phi_0/2$, and $V = \frac{\xi \hbar e}{C_G E_C \Delta t} + \frac{e}{C_G}$. Note that an overall phase shift given by $\exp(-iE_o\Delta t/\hbar)$ is also introduced. Similarly, $R_x(\xi)$ can be achieved within time $\Delta t$ by setting $\Phi_{ext} = \frac{\Phi_0}{\pi}\cos^{-1}\left(\frac{\xi \hbar}{2E_J^{intrinsic}\Delta t}\right)$, and $V = \frac{e}{C_G}$. These settings cause, within time $\Delta t$, $U_1$ to take the form $R_z(\xi) = \begin{pmatrix} e^{i\xi/2} & 0 \\ 0 & e^{-i\xi/2} \end{pmatrix}$ or $R_x(\xi) = \begin{pmatrix} \cos(\xi/2) & i\sin(\xi/2) \\ i\sin(\xi/2) & \cos(\xi/2) \end{pmatrix}$ respectively.

An overall phaseshift can be applied by setting $n_c = C_g V/2e = 1/2$. In this case $E_o = E_C/4$. The transformation is given by $Ph(\varphi) = \begin{pmatrix} e^{-i\varphi} & 0 \\ 0 & e^{-i\varphi} \end{pmatrix} = \begin{pmatrix} e^{\frac{-iE_C\Delta t}{4\hbar}} & 0 \\ 0 & e^{\frac{-iE_C\Delta t}{4\hbar}} \end{pmatrix}$.

---

′ The doubling of the angle arises because of the relationship between operations in SO(3) (rigid-body rotations) to operations in SU(2).

Thus, by the factorization given in equation (5), an arbitrary 1-qubit gate can be achieved in the SCB-based approach to quantum computing in a time of $3\Delta t$.

Note that the only free parameters used to determine the action of the 1-qubit gate are the external flux $\Phi_{ext}$ and the voltage $V$. The time interval, $\Delta t$, over which the Hamiltonian needs to act to bring about an *x*- or *z*-rotation, is fixed by the physics of the particular material, e.g., Aluminum or Niobium, used for the qubit. Although we could also have used $\Delta t$ as an additional control parameter, such a choice would complicate integration of quantum gates into parallel, synchronous, quantum circuits.

## 4    Two-Qubit Gates

To achieve a 2-qubit gate, it is necessary to couple pairs of qubits. In our scheme, two qubits are coupled using two tunnel junctions connected in parallel. This allows the coupling to be turned on or off as necessary. A schematic for the 2-qubit gate is shown in Figure 3.

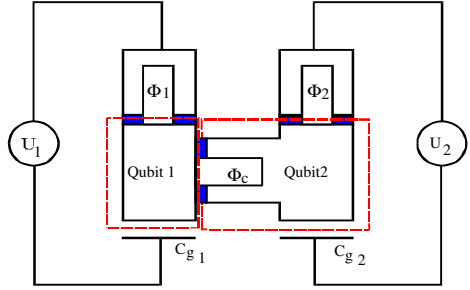

Figure 3. Schematic diagram of a pair of coupled qubits.

The Hamiltonian for the coupled pair of qubits is given by:

$$H_2 = E_{C_1}(n_1 - n_{C_1})^2 + E_{C_2}(n_2 - n_{C_2})^2 + E_{CC}(n_1 - n_2 - n_{C_1} - n_{C_2})^2 - E_{J_1}(\Phi_1)\cos(\phi_1) - \\ E_{J_2}(\Phi_2)\cos(\phi_2) - E_{J_C}(\Phi_C)\cos(\phi_1 - \phi_2) \quad (10)$$

where $E_{cc}=e^2/2C_c$, $C_c$ the capacitance of the coupling between the two qubits, the subscripts 1, 2, and *C*, refer to parameters of qubit 1, qubit 2 and the coupling between them respectively. Assuming again that the zero of energy is at $E_o=E_{c1}(n_{c1}^2-n_{c1}+1/2)+E_{c2}(n_{c2}^2-n_{c2}+1/2)^2 +E_{cc}(1-(n_{c1}-n_{c2})^2)$

$$H_2 = \begin{pmatrix} -\frac{E_1}{2}-\frac{E_2}{2}-E_{CC} & -\frac{1}{2}E_{J_2}(\Phi_2) & -\frac{1}{2}E_{J_1}(\Phi_1) & 0 \\ -\frac{1}{2}E_{J_2}(\Phi_2) & -\frac{E_1}{2}+\frac{E_2}{2}+\frac{E_{12}}{2} & -\frac{1}{2}E_{J_C}(\Phi_C) & -\frac{1}{2}E_{J_1}(\Phi_1) \\ -\frac{1}{2}E_{J_1}(\Phi_1) & -\frac{1}{2}E_{J_C}(\Phi_C) & \frac{E_1}{2}-\frac{E_2}{2}-\frac{E_{12}}{2} & -\frac{1}{2}E_{J_2}(\Phi_2) \\ 0 & -\frac{1}{2}E_{J_1}(\Phi_1) & -\frac{1}{2}E_{J_2}(\Phi_2) & \frac{E_1}{2}+\frac{E_2}{2}-E_{CC} \end{pmatrix} \quad (11)$$

where $E_1=E_{c1}(1-2n_{c1})$, $E_2=E_{c2}(1-2n_{c2})$ and $E_{12}=E_{cc}(n_{c1}-n_{c2})$. The 2-qubit quantum gate induced by this Hamiltonian is

$$U_2 = \exp(-\frac{iH_2 t}{h}) \quad (12)$$

We can specialize $U_2$ to a particular form by setting $n_{C_1} = n_{C_2} = \frac{1}{2}$, $E_{J_1} = E_{J_2} = 0$. These values induce the 2-qubit gate

$$U_2 = \begin{pmatrix} \exp\left(\frac{iE_{CC}\Delta t}{h}\right) & 0 & 0 & 0 \\ 0 & \cos\left(\frac{E_{J_C}\Delta t}{2h}\right) & i\sin\left(\frac{E_{J_C}\Delta t}{2h}\right) & 0 \\ 0 & i\sin\left(\frac{E_{J_C}\Delta t}{2h}\right) & \cos\left(\frac{E_{J_C}\Delta t}{2h}\right) & 0 \\ 0 & 0 & 0 & \exp\left(\frac{iE_{CC}\Delta t}{h}\right) \end{pmatrix} \quad (13)$$

By setting $E_{J_C} = h\pi/(2\Delta t)$ and $E_{JC} = \frac{2m+1}{n}E_{cc}$, where $m$ and $n$ are integers, we achieve a 2-qubit gate that we call the complex SWAP, $i$SWAP:

$$iSWAP = \begin{pmatrix} 1 & 0 & 0 & 0 \\ 0 & 0 & i & 0 \\ 0 & i & 0 & 0 \\ 0 & 0 & 0 & 1 \end{pmatrix} \quad (14)$$

Note that necessarily $n$ will be larger than $m$, since $E_{cc}$ is larger than the maximum value of $E_{JC}$.

## 5   Universal Quantum Computation

The set of all 1-qubit gates together with controlled-NOT is known to be universal for quantum computation. As we have already shown that it is possible to implement any 1-qubit gate in the SCB context, we can prove that all 1-qubit gates and $iSWAP$ is also a universal set by exhibiting a construction for $CNOT$ using only 1-qubit gates and $iSWAP$. The following gate sequence achieves $CNOT$:

$$CNOT \equiv \left(I_2 \otimes Ph(\frac{3\pi}{4})R_Z(\frac{\pi}{2})R_Y(\frac{\pi}{2})R_Z(\frac{\pi}{4})\right) \cdot (iSWAP \cdot iSWAP \cdot iSWAP) \cdot \\ \left(R_Z(\frac{\pi}{4})R_Y(\frac{\pi}{2})R_Z(\pi)\right) \otimes \left(R_Z(\frac{\pi}{2})\right) \cdot iSWAP \quad (12)$$

Thus a controlled-NOT operation can be implemented within the SCB-based approach to quantum computing. We should point out that an implementation of $CNOT$ using the square root of $iSWAP$ has been theoretically demonstrated in the context of quantum dots coupled via photon modes in a high Q cavity[26].

## 6   Conclusion

We have designed a realizable set of quantum gates to support universal quantum computation in the context of SCB-based quantum computing. In selecting our universal gate set we paid special attention to two principles of good computer design, namely, that each gate operation

P. Echternach, C.P. Williams, S.C. Dultz, P. Delsing, S. Braunstein, and J.P. Dowling    149should take a fixed and predictable length of time, and that the operations needed to bring about the action of a particular gate should not depend upon the time at which the gate operation begins. Earlier proposals for SCB-based universal quantum computation did not satisfy these criteria. The circuits we designed can be implemented using standard electron beam lithography and are being fabricated at the Jet Propulsion Laboratory. Tests of the RF-SET readout are already under way and will be the subject of another publication.

The research described in this publication has been carried out at the Jet Propulsion Laboratory, California Institute of Technology, under a contract with the National Aeronautics and Space Administration. This work was partially funded by the National Security Agency and the Advanced Research and Development Activity.### References

1. A. Shnirman, G. Sch n, and Z. Hermon, Phys. Rev. Lett. **79**, 2371 (1997).
2. D. V. Averin,, Solid State Communications, **105**, (1998), pp.659-664.
3. Y. Makhlin, G.Sch n, and A. Shnirman, Nature **398,** 305 (1999).
4. M. Bocko, A. Herr and M. Feldman, IEEE Trans. Appl. supercond. 7,3638 (1997).
5. J. E. Mooij, T. P. Orlando, L. Levitov, L. Tian, C. H. van der Wal, and S. Lloyd, Science, **285**, 1036 (1999).
[6] L. Tian, L. Levitov, C. H. van der Wal, J. E. Mooij, T. P. Orlando, S. Lloyd, C. Harmans, J. J. Mazo, http://xxx.lanl.gov/abs/cond-mat/9910062 (1999).
7. G. Blatter, V. Geshkenbein, and L. Ioffe, http://xxx.lanl.gov/abs/cond-mat/9912163 (1999).
8. L.B. Ioffe, V.B. Geshkenbein, M.V. Feigelman, A.L. Fauchere, G. Blatter, Nature **398**, 679 (1999).
9. A. Zagoskin, http://xxx.lanl.gov/abs/cond-mat/9903170 (1999).
10. A. Blais and A. Zagoskin, http://xxx.lanl.gov/abs/cond-mat/9905043 (1999).
11. D. Loss, D. DiVincenzo, Phys. Rev. A 57, 120 (1998).
12. D. Loss, G. Burkard, and E. V. Sukhorukov, Quantum Computing and Quantum Communication with Electrons in Nanostructures, to be published in the proceedings of the XXXIVth Rencontres de Moriond Quantum Physics at Mesoscopic Scale , held in Les Arcs, Savoie, France, January 23-30, (1999).
13. D.P. DiVincenzo, G. Burkard, D. Loss, and E. V. Sukhorukov, Quantum Computation and Spin Electronics, to be published in Quantum Mesoscopic Phenomena and Mesoscopic Devices in Microelectronics, eds. I. O. Kulik and R. Ellialtioglu, NATO Advanced Study Institute, Turkey, June 13-25, (1999).
14. R. Vrijen, E. Yablonovitch, K. Wang, H. Jiang, A. Balandin, V. Roychowdhury, T. Mor, D. Di Vincenzo, http://xxx.lanl.gov/abs/quant-ph/9905096 (1999).
15. B. Kane, Nature, 393, 133 (1998).
16. B. Kane, http://xxx.lanl.gov/abs/quant-ph/0003031 (2000). Submitted to Fortschritte der Physik Special Issue on Experimental Proposals for Quantum Computation.
17. Y. Nakamura, Yu. A. Pashkin, and J. S. Tsai, Nature **398**, 786 (1999).
18. R. Fazio, G. Massimo Palma, and J. Siewert, Phys. Rev. Lett., **83**, 5385 (1999).
19. G. Sch n, A. Shnirman, and Y, Makhlin, http://xxx.lanl.gov/abs/cond-mat/9811029 (1998). See the paragraph following their equation (7).
20. D. Di Vincenzo, Phys. Rev. A, **51**, 1015 (1995).
21. X. Miao, http://xxx.lanl.gov/abs/quant-ph/0003068 (2000).
22. X. Miao, http://xxx.lanl.gov/abs/0003113 (2000).